\documentclass[preprint2]{aastex}
\usepackage{natbib}

\def\Lyr{{\rm\,Lyr}}
\def\km{{\rm\,km}}
\def\kpc{{\rm\,kpc}}

\def\s{{\rm\,s}}

\def\g{{\rm\,g}}

\shortauthors{Koch \& Wright}
\shorttitle{Turbulent Spacetime and Dark Matter}

\begin{document}
\title{
      Can Effects of Dark Matter be Explained by the Turbulent Flow of Spacetime?
      }
\author{F. Elliott Koch and Angus H. Wright}
\affil{School of Physics, University of New South Wales, Sydney, NSW, Australia}

\begin{abstract} 
For the past forty years the search for dark matter has been one of the primary foci 
of astrophysics, although there has yet to be any direct evidence for its existence \citep{Porter2011}. 
Indirect evidence for the existence of dark matter is largely rooted in the rotational
speeds of stars within their host galaxies, where, instead of having a $\sim r^{-1/2}$ radial dependence,
stars appear to have orbital speeds independent of their distance from the galactic center, 
which led to proposed existence of dark matter \citep{Porter2011, Peebles1993}. 
We propose an alternate explanation for the observed stellar motions within galaxies, combining the 
standard treatment of a fluid-like spacetime with the possibility of a ``bulk flow'' of mass 
through the Universe. The differential ``flow'' of spacetime could generate vorticies capable
of providing the ``perceived'' rotational speeds in excess of those predicted by Newtonian mechanics.
Although a more detailed analysis of our theory is forthcoming, we find
a crude ``order of magnitude'' calculation can explain this phenomena. We also find that this can 
be used to explain the graviational lensing observed around globular clusters like ``Bullet Cluster''.
\end{abstract}  
  
\section{Introduction}  
\label{sec:Intro}  

In the pursuit of determining a model that accurately predicts the past, present, and future of the evolution 
of the Universe, physicists have generated a range of possible candidates. Currently, the most generally accepted 
being the $\Lambda$CDM model, which contains, among others, parameters dealing with the existence of a cosmological 
constant($\Lambda$) and cold dark matter (CDM). Furthermore, the existence of the dark matter component of this 
cosmology, and others, is not that which is generally considered contentious. Dark matter has rather become 
somewhat of a staple in the diet of cosmologies. However, there are observational reasons to give pause to the 
assumed existence of Universal cold dark matter, which then should lead us to question whether or not there are 
other alternative models. 

Models of dark matter succeed in accounting for the galactic rotation curves observed throughout the Universe, 
by increasing the mass of the galaxy beyond the observed. There are, however, simple problems with the dark matter 
halo model that have yet to be fully explained 
(e.g. \citep{Klypin1999, Garbari2011, Karachentsev2012, Poitras2012, Sluse2012}).
One of these problems is the disparity between the observed (stellar) 
mass function, usually defined in terms of the Schechter function, and the theoretically expected cosmological halo 
mass function \citep{Moster2010}. 
One of the defining problems of galactic formation and evolution is determining the origins of this 
disparity. This is an example of how our understanding of dark matter (or lack there of) is still grounds for much 
debate. However, if it may be possible to utilise a different model for the origin of galactic formation then it is 
possible that some of these questions may be answered.

Recently, there has been some observational evidence for the ``bulk flow'' of matter through the Universe
\citep{Benson2003, Bhattacharya2008, Feldman2010, Osborne2011, Turnbull2012}. 
If these measurements prove to be true, then the nature of this flow is of interest beyond that of the
distribution of matter in the Universe. Specifically relevant to this 
discussion is the interpretation that the ``flow'' observed is caused not by an {\em en masse} transit of matter 
through the Universe, but rather by the motion of spacetime itself. Whilst this concept is 
indeed foreign it can be considered somewhat preferential to the former case, from an isotropic 
viewpoint, as the motions of objects in the Universe need not be preferentially oriented in this regime. 
More importantly, variation in the ``bulk flow'' of spacetime fluid through the Universe could produce eddies 
in spacetime and provide the additional unexplained velocity to rotational speeds of stars beyond the 
central bulge of galaxies. 

As the intention of this paper is to merely propose an alternative theoretical explanation for observations
consistent with the existence of dark matter the structure is as follows: \S~\ref{sec:spacetime} describes
the treatment of spacetime as a fluid. Sections \ref{sec:classical} and \ref{sec:relativistic} discuss
classical fluid dynamics and relativistic fluid dynamics and how votices in the spacetime ``fluid'' can
produce observations consisten with dark matter. Finally we present concluding remarks and propose future
work in \S~\ref{sec:summary}.

\section{Spacetime as a Fluid and the Differential Rotation of Spacetime}
\label{sec:spacetime}
In General Relativity it is common for theories to compare the nature of the spacetime coordinate 
system to an ideal fluid, often called the `cosmic fluid'. This treatment is integral for the 
formulation of many concepts in GR, including the formation and propagation of gravitational waves. 
Additionally, Kerr-Newman geometry for a rotating black hole, which predicts 
an effective ``differencial rotation'' of spacetime. 
The principle effect regarding black hole studies is the 
implication that, within $ r_0 \equiv M + \sqrt{M^2 - a^2 \cos^2\theta}$, it is impossible to remain stationary
with respect to distant ``stationary'' observer beyond $r_0$. 

The Kerr-Newman geometery has been used to successfully describe the light curves of rapidly
rotating neutron stars \citep{Braje2000}. Doppler boosting and time-delay-induced pertubations 
from frame-dragging cause ``soft lags'' in pulse profiles of neutron stars which have been measured
in X-ray spectra.

As a starting point for our proposition 
we use this geometry, simply as an example of how the differential rotation of spacetime is 
possible. We leave all other black-hole allegories or implications of a large, dense,
rotating mass located at the ``centre of the Universe'' behind. 

\section{Vortices in Classical Fluid Dynamics}
\label{sec:classical}
If spacetime is able to experience a form of differential rotation, then it is of interest to examine 
how this would impact the fluid treatment we noted earlier. 

Classical fluid dynamics states that eddies/vortices with angular
velocity $\omega^{\rm vort}$ occur for any differential flow as: 
\begin{equation}
 \omega^{\rm vort} = \nabla \times \mathbf{v} 
 \label{eq:ClasVort}
\end{equation}
 and 
\begin{displaymath}
 \frac{\partial \omega}{\partial t}^{\rm vort} = \nabla \times (\omega^{\rm vort} \times \mathbf{v}).
\end{displaymath}
Due to the chaotic nature of this relationship, when applied to astronomical distances and timescales 
even small perturbations can ultimately produce large scale phenomena. 

Essentially, if we treat spacetime as a classical fluid with turbulence 
caused by a differential flow as observed by \citet{Osborne2011} (and others), 
spatial variations in the flow would generate 
eddies that could not only host galaxies, but adequately explain their rotational dynamics as well. 

Treating spacetime as a classical fluid, and assuming the velocity of any ``bulk flow'' is
$\mathbf{v} = v_x(x,y) \hat{x} + v_y(x,y) \hat{y}$,  we find
$ \omega^{\rm vort}_z = ( \delta v_x^2 + \delta v_y^2 )^{1/2}$ where 
$\delta v_i = \partial v_i / \partial x_{j \ne i}$. \citet{vanAlbada1986} utilised light 
curves to show that rotational velocities of stars deviated from Newtonian mechanics by $\sim 50$ km/s
at a radial distance of $\sim 20$ kpc from the galactic centre. 
We find that, to account for this discrepancy $\Delta v \equiv ( \delta v_x^2 + \delta v_y^2 )^{1/2}$,  
$\omega^{\rm vort} = \Delta v \sim 3 \km \s^{-1} \kpc^{-1}$ for observed velocities of the 
outer most stars of a galaxy which correspond to the largest values of 
$\Delta v$. So, given
some bulk flow through in the local Universe, there only need be a systematic variation
of $\sim 3 \km \s^{-1} \kpc^{-1}$ within the flow to produce an eddy large enough to provide
the unaccounted velocity to the outermost stars of a galaxy. Furthermore, as the velocity of the eddy is 
dependent on radius from the centre of the eddy (i.e. the galactic centre), the effect will diminish 
as the radius from the galactic centre decreases. Accounting for the increase in observable galactic mass 
with decreasing radius, it can be supposed that as one approaches the galactic centre the motion of stars 
becomes primarily governed by gravitation. Formulated simply: as $r \rightarrow r_c$, $v_{\star} \rightarrow v_{N}$, 
where $r_c$ is the radius of the galactic ``bulge'' and $v_N$ is the orbital speed of stars predicted by 
Newtonian mechanics.


In this way, we propose that the observed motion of galactic stars 
over the entire disk may be explained by the presence of 
eddies in spacetime caused by appreciably small variations in ``bulk flow''. 

Thus far we have treated spacetime as a classical fluid, but as we are dealing with
distortions of spacetime relativistic effects must be addressed. Greenberg found
that, independent of any relativistic geometry used, the angular velocity 2-form of an eddy/vortex is 
\begin{equation}
 \omega_{\alpha \beta} = \frac{1}{2} \left(u_{\alpha ; \beta} - u_{\beta ; \alpha} \right)
 \label{eq:RelVort}
\end{equation}
where $u_{\beta}$ is the velocity 4-vector of the ``fluid'' \citep{MTW}.
The most noticeable difference between equations (\ref{eq:ClasVort}) and (\ref{eq:RelVort}) is
the coupled nature of spacetime. Therefore any changes in the ``flow'' over time could also produce 
vortices in spacetime, much like those produced by variations in differential flow in classical
dynamics. Coupling temporal and spatial variations could further enhance the turbulent
nature of spacetime, and thus the production of eddies where
galaxies could grow.

\section{Relativis Fluid Dynamics}
\label{sec:relativistic}
Examining the relativistic case, we can choose a reference frame such that $ u_t \ne 0$, and
the spatial components $u_i$ are all zero. The reasoning behind this is that, for an external
observer, the motion of a ``fluid'' through a stationary reference frame is observationally indifferent
to a static ``fluid'' in a co-moving reference frame. Using this restriction, equation (\ref{eq:RelVort})
becomes
\begin{equation}
 \omega_{ti} = -\omega_{it} = \frac{1}{2} u_{t ; i} \\
\end{equation}
Since this essentially a 2-form for a vortex in spacetime, the magnitude of the resulting rotation
would be 
\begin{eqnarray}
 \omega^2 &=& \omega_{\alpha \beta}\omega^{\alpha \beta} \nonumber \\
          &=& \omega_{\alpha \beta} g^{\alpha \gamma} g^{\beta \delta} \omega_{\gamma \delta},
 \label{eq:omega2}
\end{eqnarray}
which is the magnitude of the rotation for the vortex, squared. Since it contains products of 
$\omega_{\mu\nu} \omega_{\nu\mu}$, then, in principle, $\omega^2$ can be non-zero. 
Additionally, the covariant derivative
embedded in $\omega_{\mu\nu}$ provides us with information on variations within $\g_{\mu\nu}$ (spacetime)
that yield $\omega^2 \ne 0$, possibly explaining observed phenomena. 

We reserve the comprehensive mathematical analysis for a forthcoming paper and merely present 
a simplified ``glimpse'' of the interpretation of equation (\ref{eq:omega2}). If a vortex in spacetime
is to be used as a possible explanation for the observed rotational velocities of stars, and
on average, these velocities are independent of angular position within the galaxy as 
well as vertical position within the disc, we have assumed that $u_{t ; \phi} = u_{t ; z} = 0$, 
hence 
\begin{eqnarray}
 \omega^2 &=& 2 \omega^2_{rt} \left[g^{tt}g^{rr} - \left(g^{tr}\right)^2\right] \nonumber \\
          &=& \frac{1}{4} \left(\partial_r u_t - \Gamma^t_{rt} u_t \right)^2 
             \left[g^{tt}g^{rr} - \left(g^{tr}\right)^2\right],
\end{eqnarray}
because $g^{tr}=g^{rt}$ and $\omega_{tr} = -\omega_{rt}$. This can be used as a restriction
for properties of $g_{\mu \nu}$ and expand upon existing geometries such as the Kerr-Newman and/or
Friedmann-Robertson-Walker metrics, which will be presented in subsequent publications. Here
we merely present evidence that small perturbations in $g_{\mu \nu}$ and $u_t$ may be used
to explain the rotational speeds of stars in other galaxies.

Assuming the metric for the local spacetime is essentially flat, 
$g_{\mu \nu} = \delta_{\mu \nu} + \delta g_{\mu \nu}$, as well as the variations in $u_t$ 
$\delta u_t \equiv \partial_r u_t$ and the variations in $g$ as $\delta g u_t \equiv \Gamma^t_{rt} u_t$,
we can approximate the magnitudes of these ``variations'' that could explain observed phenomena.
Since we are assuming cross terms are small and $g^{tt} \sim g^{rr} \sim 1$, 
$\omega^2 \sim (\delta u_t - \delta g u_t)^2$. In the extreme case where the rotational speed
of stars for a galaxy, as predicted by theory is zero, $v_{\star} = \omega r_{\S}$ or
$ \omega = v_{\star}/r_{\S} \sim 10^{-4} \km \s^{-1} \Lyr^{-1}$ and 
that $\delta u_t = 0$, implying that the stellar motion is purely from variations in the 
geometry. Coupling this with the afore mentioned indifference of motion through spacetime and 
spacetime moving with the ``fluid'' we can approximate the magnitude of $\delta g$. 
Based on observations of \citet{Osborne2011}, we assume that $u_t \sim 100 \km \s^{-1}$.
Therefore, we determine that $\delta g \sim 10^{-6} \Lyr^{-1}$ could produce effects consistent 
with stellar rotational velocities observed. Additionally, \citet{Osborne2011}  also found that
the ``flow'' changed by $\sim 50 \km \s^{-1}$ between the redshifts of 0.4 and 0.8.
Again, if this is largely due to variations in the geometry of spacetime, this would imply
$\delta g \sim 10^{-8} \km \s^{-1}$. Though this is smaller than the previous approximation, 
that approximation did not account for the motion of stars from Newtonian mechanics.

\section{Summary}
\label{sec:summary}
If the observed effects attributed to dark matter are indeed caused by the turbulent flow of spacetime, then
we can simply hypothesise that any galaxy in a cluster formed in this way should all rotate in the same 
direction. Albeit only significant to $1.6 \sigma$, evidence for this effect has been detected by 
\citet{Longo2011}. 
There is no reason to expect that this observation would also be caused in the dark matter Regime for reasons 
other than chance. As this is a simple method to determine
if turbulent spacetime flows may be the cause of galactic rotational curves, Doppler
measurements of stellar velocities in galaxies are extremely important. Furthermore, if rotational curves for 
distant galaxies can be found, isotropy measurements could serve as an additional constraint for the
validity of this theory. 

Finally, we find that chaotic flows could exist on both ``large'' and ``small'' scales. Large scale turbulence 
is dominated by the ``flow'' velocity and the uniformity thereof. In this context, ``large'' scale turbulence 
would be on a galactic scale, with ``large'' eddies being comparable to the size of a galaxy, which
could be used to explain why galaxies are not ``sheared'' apart. Small scale 
turbulence is dominated by the viscosity of the flow, which would most likely be 
caused by the gravitational attraction of masses present in the region of the eddy, as mentioned previously 
in relation to the galactic rotation curves. The scale of a ``small
eddy'' would be comparable to that of stellar clusters. Since these clusters are still affected by
distortions in spacetime, this could explain the observed gravitational lensing caused by some
stellar clusters that could not be explained by modified Newtonian dynamics (MOND) \citep{Ibata2011}. 

This concept can, therefore, effectively explain the major observations that lead to the introduction of 
dark matter, and removes the need for the existence of a massive dark matter halo about galaxies. 
If this theory is correct, galaxies co-located within a differentially moving frame, should all rotate 
in the same direction (i.e. same chirality). Additionally, when observed from an external reference frame,
there should be variations in the local spacetime of a galaxy $\delta g u_t \sim 10^{-4} \km \s^{-1} \Lyr^{-1}$
where $u_t$ is the observed ``flow'' of the surrounding galaxies.

Finally can now draw some different, interesting, conclusions from
the disparity between the observed stellar mass function and the cosmological halo mass function. 
Using the model we propose here, the origins of galactic evolution 
lie not in vast halos of dark matter, but rather in the turbulence of spacetime. The turbulence itself traces back its 
origins to spatial and temporal variations in the motion of spacetime. If we are able to conceptualise this method of 
seeding galaxies, then we can also recognise that the formation of the vortices in which galaxies originate is 
highly dependent on the mechanics of the local spacetime. This, when interpreted simply, means that the expected 
distribution of galactic mass (i.e. the observed stellar mass function) should not be conformal to a simplistic power 
law (i.e. the cosmological halo mass function), but rather should be more complex in nature. The true distribution of 
vortex sizes in a field of uniform variation should be expected to be inately coupled with galactic mass.
 This may well explain the shape of the Schechter function, and why galactic mass function does not follow 
a simple power law, without the need for complex models to explain the reduction of star formation in the low and high 
mass dark matter halo regimes. 

As stated above, many of these conclusion are too complicated to present in any detail within this paper, of which
will be presented in following papers. The purpose of this paper is to merely propose an alternate explanation
for the observed effects of dark matter and a possible explanation for why dark matter is not consistent with some other 
theories and observation 

\section{Acknowledgements}
The authors would like to thank Graeme Salter, Jonathan Horner, Annant Tanna and Bradley Hansen for insightful
discussions regarding the formulation of this theory. FEK would also like to thank UNSW and the ARC for financial
support used during the development of this theory under grants PS27228-FRGP, PS27261-ECR and DP120105045.

\bibliographystyle{apj}
\bibliography{thesis}

\begin{thebibliography}{18}
\expandafter\ifx\csname natexlab\endcsname\relax\def\natexlab#1{#1}\fi

\bibitem[{{Benson} {et~al.}(2003){Benson}, {Church}, {Ade}, {Bock}, {Ganga},
  {Hinderks}, {Mauskopf}, {Philhour}, {Runyan}, \& {Thompson}}]{Benson2003}
{Benson}, B.~A., {Church}, S.~E., {Ade}, P.~A.~R., {Bock}, J.~J., {Ganga},
  K.~M., {Hinderks}, J.~R., {Mauskopf}, P.~D., {Philhour}, B., {Runyan}, M.~C.,
  \& {Thompson}, K.~L. 2003, \apj, 592, 674

\bibitem[{{Bhattacharya} \& {Kosowsky}(2008)}]{Bhattacharya2008}
{Bhattacharya}, S. \& {Kosowsky}, A. 2008, \prd, 77, 083004

\bibitem[{{Braje} {et~al.}(2000){Braje}, {Romani}, \& {Rauch}}]{Braje2000}
{Braje}, T.~M., {Romani}, R.~W., \& {Rauch}, K.~P. 2000, \apj, 531, 447

\bibitem[{{Feldman} {et~al.}(2010){Feldman}, {Watkins}, \&
  {Hudson}}]{Feldman2010}
{Feldman}, H.~A., {Watkins}, R., \& {Hudson}, M.~J. 2010, \mnras, 407, 2328

\bibitem[{{Garbari} {et~al.}(2011){Garbari}, {Read}, \& {Lake}}]{Garbari2011}
{Garbari}, S., {Read}, J.~I., \& {Lake}, G. 2011, \mnras, 416, 2318

\bibitem[{{Ibata} {et~al.}(2011){Ibata}, {Sollima}, {Nipoti}, {Bellazzini},
  {Chapman}, \& {Dalessandro}}]{Ibata2011}
{Ibata}, R., {Sollima}, A., {Nipoti}, C., {Bellazzini}, M., {Chapman}, S.~C.,
  \& {Dalessandro}, E. 2011, \apj, 738, 186

\bibitem[{{Karachentsev}(2012)}]{Karachentsev2012}
{Karachentsev}, I.~D. 2012, Astrophysical Bulletin, 67, 123

\bibitem[{{Klypin} {et~al.}(1999){Klypin}, {Kravtsov}, {Valenzuela}, \&
  {Prada}}]{Klypin1999}
{Klypin}, A., {Kravtsov}, A.~V., {Valenzuela}, O., \& {Prada}, F. 1999, \apj,
  522, 82

\bibitem[{{Longo}(2011)}]{Longo2011}
{Longo}, M.~J. 2011, Physics Letters B, 699, 224

\bibitem[{{Misner} {et~al.}(1973){Misner}, {Thorne}, \& {Wheeler}}]{MTW}
{Misner}, C.~W., {Thorne}, K.~S., \& {Wheeler}, J.~A. 1973, {Gravitation} (San
  Francisco: W.H.~Freeman and Co., 1973)

\bibitem[{{Moster} {et~al.}(2010){Moster}, {Somerville}, {Maulbetsch}, {van den
  Bosch}, {Macci{\`o}}, {Naab}, \& {Oser}}]{Moster2010}
{Moster}, B.~P., {Somerville}, R.~S., {Maulbetsch}, C., {van den Bosch}, F.~C.,
  {Macci{\`o}}, A.~V., {Naab}, T., \& {Oser}, L. 2010, \apj, 710, 903

\bibitem[{{Osborne} {et~al.}(2011){Osborne}, {Mak}, {Church}, \&
  {Pierpaoli}}]{Osborne2011}
{Osborne}, S.~J., {Mak}, D.~S.~Y., {Church}, S.~E., \& {Pierpaoli}, E. 2011,
  \apj, 737, 98

\bibitem[{{Peebles}(1993)}]{Peebles1993}
{Peebles}, P.~J.~E. 1993, {Principles of Physical Cosmology}

\bibitem[{{Poitras}(2012)}]{Poitras2012}
{Poitras}, V. 2012, \jcap, 6, 39

\bibitem[{{Porter} {et~al.}(2011){Porter}, {Johnson}, \& {Graham}}]{Porter2011}
{Porter}, T.~A., {Johnson}, R.~P., \& {Graham}, P.~W. 2011, \araa, 49, 155

\bibitem[{{Sluse} {et~al.}(2012){Sluse}, {Chantry}, {Magain}, {Courbin}, \&
  {Meylan}}]{Sluse2012}
{Sluse}, D., {Chantry}, V., {Magain}, P., {Courbin}, F., \& {Meylan}, G. 2012,
  \aap, 538, A99

\bibitem[{{Turnbull} {et~al.}(2012){Turnbull}, {Hudson}, {Feldman}, {Hicken},
  {Kirshner}, \& {Watkins}}]{Turnbull2012}
{Turnbull}, S.~J., {Hudson}, M.~J., {Feldman}, H.~A., {Hicken}, M., {Kirshner},
  R.~P., \& {Watkins}, R. 2012, \mnras, 420, 447

\bibitem[{{van Albada} \& {Sancisi}(1986)}]{vanAlbada1986}
{van Albada}, T.~S. \& {Sancisi}, R. 1986, Royal Society of London
  Philosophical Transactions Series A, 320, 447

\end{thebibliography}

\label{lastpage}
\end{document}